\documentclass{appolb}
\usepackage{amssymb}
\usepackage{epsfig,dcolumn}
\usepackage{graphicx}
\usepackage{graphicx}
\usepackage{bm}
\usepackage{epsfig,dcolumn}
\usepackage[usenames]{color}

\newcommand{\ar}{\arrowvert}
\newcommand{\ra}{\rangle}
\newcommand{\la}{\langle}
\newcommand{\da}{\dagger}

\newcommand{\cd}{\! \cdot \!}
\newcommand{\be}{\begin{equation}}
\newcommand{\ee}{\end{equation}}
\newcommand{\bea}{\begin{eqnarray}}
\newcommand{\eea}{\end{eqnarray}}
\begin{document}


\title{Excited baryons as experimental probes of the quark mass}


\author{
Felipe J. Llanes-Estrada
\address{
Universidad Complutense de Madrid, Spain}
\and
Pedro Bicudo, Marco Cardoso
\address{Instituto Superior T{\'e}cnico, Lisboa, Portugal}
\and
Tim van Cauteren
\address{Dept. Sub. and Rad. Phys., Ghent University, Belgium}
}

\maketitle

\begin{abstract}
We observe that excited hadrons provide an opportunity to probe from experiment the power-law running of the quark mass in the mid infrared, while the condition $m(k)<k$ remains valid. \\
A relatively clean analysis is possible for the maximum spin 
excitations of the $\Delta$ baryons, analogous to the yrast states in 
nuclei. Such states are accessible at current experimental 
facilities such as ELSA and Jlab.
\end{abstract}
\PACS{ 
 11.10.St,  11.30.Rd, 14.20.-c, 14.65.Bt
}


Quark masses are fundamental parameters of the Standard Model that appear in the  Lagrangian density of Quantum Chromodynamics, whose non-interacting part reads
\be
{\mathcal{L}}_{\rm{QCD}}^0 = \sum_i m_i \bar{\psi} \psi  + i \sum_i
\bar{\psi} \not \partial \psi .
\ee
These masses can also be read off an expansion of the quark propagator for large momenta in perturbation theory
\be
S(k) = \frac{i}{\not k - m} 
\ee
and information about them can be gathered in several ways. For light quarks, only very coarse information is available from sum rules and lattice gauge theory, although quark mass ratios are somewhat better known from chiral perturbation theory~\cite{Amsler:2008zzb}.

As the momentum of the propagating quarks becomes smaller towards the QCD scale $\Lambda_{\rm QCD}$, the mass in the denominator runs like a power-law
\be
S(p) = \frac{iZ(p^2)}{\not k - m(k^2)} \ \ \ {\rm with}\ \ \ m(k^2)\propto k^\alpha
\ee
as has been shown by Dyson-Schwinger studies~\cite{Alkofer:2003jj} in Landau gauge and lattice gauge theory~\cite{Bowman:2006zk}. For even lower $k$ the running mass function $m(k)$ presumably stabilizes and gives sense to the concept of constituent quarks, but to date there is no precise dynamical meaning to the quark masses obtained from quark model fits to data in terms of the QCD parameters.

What we have recently proposed, and elaborate in this article, is that the exponent of the power-law running in the mid-infrared is accessible to experiment by studying excited hadrons. The key idea is to exploit the recently understood parity doubling in the high hadron spectrum, by which excited hadrons decouple from the pion and instead become degenerate with a partner of equal parity, the mass difference being of order the quark mass that wants to be extracted, $|M_+-M_-|\propto m(\la k\ra)$.
The scale at which this mass is probed is the average momentum of the quark in the excited hadron. The proportionality constant is difficult to calculate, so the absolute normalization of the quark mass is not readily accessible, but by comparing increasingly excited hadrons with larger and larger $\la k\ra $, one can extract its running.

\begin{figure}[ht!]
\centerline{\includegraphics[height=2.1in]{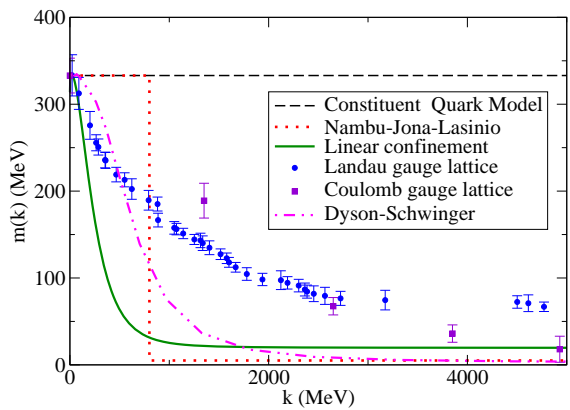}}
\caption{IR enhacement of the light quark mass, generated when spontaneous chiral symmetry breaking occurs. 
Shown are quark masses in the main approaches to QCD, 
all multiplied by an arbitrary factor to match
them at zero momentum. t
\label{fig:massgeneration}}
\end{figure}

In Hamiltonian dynamics, the running $m(k)$ defined from the propagator appears in the Dirac spinors $U_{\kappa\lambda}$.
For light quarks one can expand spinors in the ultrarelativistic  (large-momentum) limit  in the inverse ratio $m(k)/k$
\bea \!  \!  \!  \!  \!
U_{\kappa\lambda} = \frac{1}{\sqrt{2E(k)}} \left[ \begin{array}{c}
    \sqrt{E(k)+ m(k) }\chi_\lambda
\\
\sqrt{E(k)- m(k)}
    \vec{\sigma}\cdot\hat{\kappa} \chi_\lambda\end{array} \right]  \\ \nonumber 
\mathop {\longrightarrow }\limits_{k \to \infty }
 \frac{1}{\sqrt{2}} \left[ \begin{array}{c}
  \chi_\lambda \\
    \vec{\sigma}\cdot\hat{\kappa} \chi_\lambda
\end{array} \right]
+
 \frac{1}{2 \sqrt{2}} {m(k) \over k} \left[ \begin{array}{c}
 \chi_\lambda \\
-
    \vec{\sigma}\cdot\hat{\kappa} \chi_\lambda
\end{array} \right] \ \ \ \  
\label{msk_expansion}
\eea
with $E(k)=\sqrt{k^2+m(k)^2}$, 
having kept the leading chiral invariant term, and
a next order  chiral symmetry breaking $m(k) \over k$  term.
Non-chiral, spin-dependent potentials in the quark-quark interaction originate from the second term in the expansion Eq.~(\ref{msk_expansion}).

This translates into an expansion of $H^{QCD}$ in the weak sense (that is, not of the
Hamiltonian operator itself, but its restriction to the Hilbert
space of highly excited resonances, where $\langle k\rangle$ is large)
is:
\be \label{QCDexp}
\la n \ar H^{QCD} \ar n' \ra \simeq \la n \ar H^{QCD}_\chi \ar n'\ra +  \la n \ar \frac{m(k)}{k} H^{QCD\ '}_\chi \ar n' \ra + \dots
\ee

\begin{figure}[t!]
\centerline{\includegraphics[width=3in,height=1.9in]{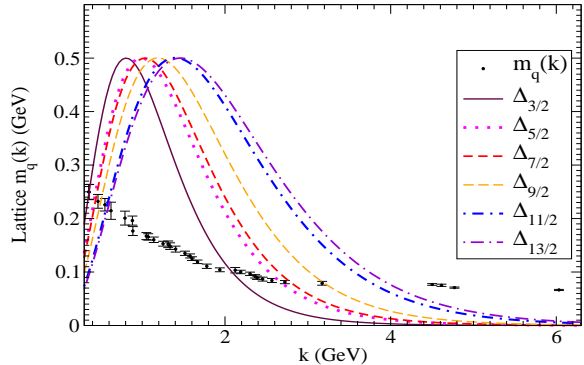}}
\caption{
Typical momentum distributions of increasingly excited 
$\Delta_{3/2}$, \dots $\Delta_{13/2}$ resonances overlap less and less with the dynamically generated IR quark mass so that $\la n\ar \frac{m(k)}{k} H^{QCD\ '}_\chi \ar n \ra$ becomes small. (Illustrative variational wavefunctions for a linear potential with string tension $\sigma=0.135$ GeV$^2$, not normalized for visibility).
This computation has been reported in~\cite{Bicudo:2009cr}.
\label{fig:anglinmoment}}
\end{figure}

The parity degeneracy~\cite{Detar:1988kn}  for highly excited hadrons follows from
invariance under transformations generated by the
chiral charge~\cite{Nefediev:2008dv}
$
Q_5^a=\int d{\bf x} \psi^\dagger (x) \gamma_5 \frac{\tau^a}{2} \psi (x)
$
and $[Q_5^a,H]=0$.
 For low--lying hadrons, Chiral Symmetry is spontaneously broken by the ground state,
$Q_5^a \ar 0 \ra \not = 0$ providing a large quark
mass in the  propagator, $m(k)$, pseudo-Goldstone bosons
($\pi,\ K,\ \eta$), and the loss
of parity degeneracy in  ground state baryons.
Substituting the spinors, and  in terms of Bogoliubov-rotated $q\bar{q}$ 
normal modes $B$, $D$, 
\begin{eqnarray} \label{chiralcharge}
Q_5^a  = \int \frac{d^3k}{(2\pi)^3} \sum_{\lambda
\lambda ' f f'c} \left(  \frac{\tau^a}{2} \right)_{ff'}
{ k \over \sqrt{ k^2 + m^2(k)}}
 \\ \nonumber 
\left( ({\bf \sigma}\cd{\bf
\hat{k}})_{\lambda \lambda'}  	
\left( B^\da_{k \lambda f c} B_{k \lambda' f' c} + D^\da_{-k \lambda' f' c}
D_{-k \lambda f c}
\right) + \right. \\  \nonumber \left.
{ m(k) \over k} (i\sigma_2)_{\lambda \lambda'} \
\left( B^{\da}_{k\lambda f c} D^\da_{-k\lambda'f'c}+
B_{k \lambda' f' c} D_{-k \lambda f c}
\right) \right)  
\end{eqnarray}
In the presence of Spontaneous $\chi$SB,
$ m(k) \not=0$,
and the last term realizes chiral symmetry non-linearly in the spectrum as it creates/destroys a pion.\\
But when $\la k \ra$ is large, it is the ${\bf \sigma}\cd{\bf
\hat{k}}$-term that dominates, and chiral symmetry is realized linearly (with only quark counting operators flipping parity and spin). 
This happens for a baryon resonance  high enough in the spectrum, 
whose constituents have a momentum distribution peaked higher than the 
IR enhancement of $m(k)$ (see figure \ref{fig:anglinmoment}.

To exploit the smallness of the $| M^{P=+} - M^{P=-}| $ mass difference, we
employ  increasing angular momentum 
and establish how the $j$-scaling of the splitting is related to the $k$-scaling of $m(k)$.\\
The $M^\pm$  in our proposed study are the masses of the ground state quartets of the $\Delta$ spectrum, with parity $\pm$ and  in the limit of large total angular momentum $j>>3/2$.
These states of maximum spin for a given energy may conveniently be called hadron {\emph{yrast}} states, in analogy with their nuclear physics counterparts (the {\emph{dizziest}} states, from the swedish {\emph{yr}}, dizzy).
They have the advantage of being {\emph{cold}}, that is, all the excitation energy is spent in rotation and no radial nor, specially, multiquark excitations are expected to be present. 

For each spin, the two quasi-degenerate masses $M^+$
and $M^-$ fall in the same (leading)  Regge trajectory, and this fixes  their $j$-scaling
\be \label{trajectory}
j =   \alpha_0 + \alpha {M^\pm}^2 
\mathop {\longrightarrow }\limits_{j \to \infty }
\alpha {M^\pm}^2 \ .
\ee
The parity of the ground state (the lowest of the two states) alternates  between $+$ an $-$
as the angular momentum steps up by one.
Large $j$  is equivalent to large quark orbital angular momentum $l$
since the spin is finite, and also to a large average linear momentum  
$\langle  k \rangle$. This is illustrated in Fig
\ref{fig:anglinmoment} where we show how the overlap of the
wavefunction with the running $m(k)$ blinds  high-lying states to $\chi$SB.
From the relativistic version of the virial theorem~\cite{virial} between kinetic and total energies, 
\be \label{Regge}
\langle  k \rangle  \to    c_2 \, M^\pm    \to {c_2 \over  \sqrt \alpha}   \sqrt j  \,
\ee 
(where $ c _2$ is a constant; for a linear potential model, and 3 quarks in a baryon, $c_2=1/6$).

The first term in Eq. (\ref{QCDexp}) cancels out in the difference
$ \ar M^+ - M^-\ar  <<  M^\pm  $
(while both $M^\pm$ are dominated by 
the chiral invariant term, their difference stems from the dynamically generated quark mass) thus exposing the second term in eq.(\ref{QCDexp}), proportional to
$ \langle  {m(k)  \over k} \rangle$, viz.
\bea \label{paritysplitofm}\! \! \!
 \ar M^+\! -\! M^-  \ar \! \to \!  \langle { m( k  )\over  k  }
H^{QCD\ '}_{\chi} 
 \rangle  \to   c_3{ m ( \langle  k \rangle ) \over  \langle  k \rangle }
\langle  H^{QCD\ '}_{\chi} \rangle   
\eea
(the factorization is allowed by the mean-value theorem at the price of a constant that we do not attempt to determine here). 
This equation is analogous to the renowned Gell-Mann-Oakes-Renner relation
$
M_\pi^2 = - m_q \frac{\la \bar{\psi} \psi \ra}{f_\pi^2}
$
but active when chiral symmetry is realized linearly, as in the high-baryon excitations we examine.

The ${\bf \sigma}\cd {\bf \hat{k}}$ from the spinors induce spin-dependent potentials in  $H^{QCD\ '}_{\chi}$.
We need its large $j$-scaling,
equivalent to large $\langle  k \rangle $ and $M^\pm$, and
separately consider its angular and radial dependences  $\langle
H^{QCD\ '}_{\chi}\rangle \; \propto \; \langle H^{QCD\ '}_{\chi}
\rangle_{\mbox{ang}} \times  \langle H^{QCD\ '}_{\chi}
\rangle_{\mbox{rad}}$. 
The angular matrix element generally includes  a spin-orbit term, that leads its $j$-counting 
\be \label{angpart}
\langle 
H^{QCD\ '}_{\chi}
\rangle_{\mbox{ang}}
 \to j 
\ee
while the centrifugal barrier dominates the radial part,  
$$
l(l+1) \langle H^{QCD\ '}_{\chi} \rangle_{\mbox{rad}} \to 
j^2    \langle H^{QCD\ '}_{\chi} \rangle_{\mbox{rad}} \to 
c_4 M^\pm
$$
and thus  $\langle H^{QCD\ '}_{\chi} \rangle_{\mbox{rad}}\propto j^{-2} M^\pm$.
From Eqs.~(\ref{trajectory},\ref{angpart}),
\be
\langle 
H^{QCD\ '}_{\chi}
\rangle
 \to  c_5 \, M^\pm  j^{-1} \to {c_5 \over  \sqrt \alpha}   \sqrt {1 \over j}  \ .
\label{tensor}
\ee
(As a corollary, note that a spin-independent potential scales like $ 1 \over j ^{ 3 / 2 }$).

The result of the $j$-scaling analysis reads then
\bea \! \!
\ar M^+ - M^-\ar\! 
 \to \! { c_3 \, m ( \langle  k \rangle )\over  \langle  k \rangle }
 \times c_5 \, M^\pm  \times j ^{-1} \! =\! 
{c_3 c_5 \over c_2 } m ( \langle  k \rangle ) \!   j ^{-1}   .
\eea
that links the IR enhancement of the quark
mass to baryon spectroscopy in a usable way. 
An experimental extraction proceeds by just
fitting the exponent of the $j$-scaling for the splitting $\ar M^+ - M^-  \ar \propto j^{-i} $.
Then, in view of eq. (\ref{Regge}), one obtains both
\bea
m(\Lambda\times \sqrt{j}) \propto j^{-i+1} \ {\rm and}\
m(k) \propto k^{-2i+2} \ .
\eea
The same exponent $i$  in this last equation can be obtained from the fit to the $\ar M^+ - M^-  \ar$ with increasing $j$.

In conclusion, the scaling with $j$ of the mass--splittings in the $\Delta$ spectrum can be related to the $k$--running of the quark mass as defined from the quark propagator. This remarkable result holds as long as the expansion in $m(\la k\ra)/\la k\ra $ makes sense, that is, for highly excited resonances.

Other consequences of Insensitivity to Chiral Symmetry Breaking high in the spectrum, such as the decoupling of excited resonances from pions (for example, the widths
$\Gamma_{\Delta^*\to N\pi}$) can likewise be exploited to probe the running quark mass.

{\emph{
We would like to remark the very pleasant and intellectually 
stimulating atmosphere of the meeting in Zakopane.
We thank L. Glozman for useful conversations, and grants
FPA 2008-00592/FPA, FIS2008-01323, CERN/FP /83582/2008, POCI/FP /81933/2007, /81913/2007 , PDCT/FP /63907/2005 and /63923/2005, Spain-Portugal billateral grant HP2006-0018 / E-56/07, as well as the Fund for Scientific
Research Flanders.
}}


\end{document}